**Non-Associativity of Lorentz Transformation and Associative Reflection Symmetric Transformation**

Mushfiq Ahmad
Department of Physics, Rajshahi University, Rajshahi, Bangladesh.
E-mail: mushfiqahmad@ru.ac.bd

M. Shah Alam
Department of Physics, Shah Jalal University of Science and Technology, Sylhet, Bangladesh.
E-mail: salam@sust.edu

## Abstract

Each of the two moving observers observes the relative velocity of the other. The two velocities should be equal and opposite. We have shown that this relativistic requirement is not fulfilled by Lorentz transformation. We have also shown that the reason is that Lorentz transformation is not associative. Reciprocal symmetric transformation is associative and fulfills relativistic requirements.

## 1. Introduction

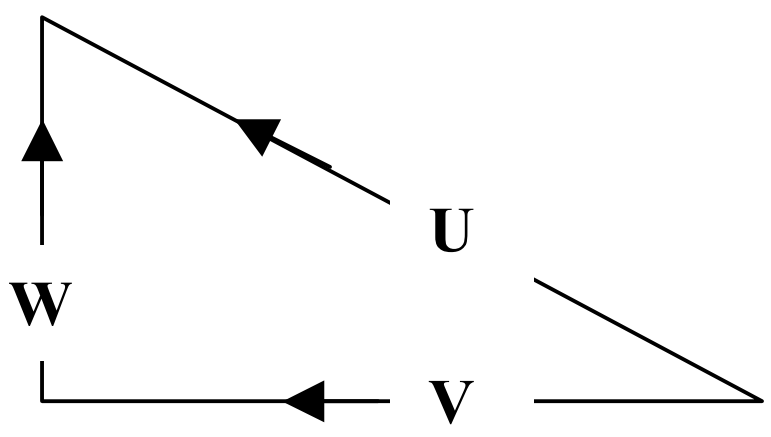

Two observers are moving with velocities **U** and **V.** Their relativity velocity according to one observer should be **W.** According to the other observer it should be –**W**. Arrows give the positive directions of the vectors. Negative vectors have the opposite directions. According to triangular law of addition of vectors, we must have $\mathbf{W} = (-\mathbf{V}) \hat{+} \mathbf{U}$ and $-\mathbf{W} = (-\mathbf{U}) \hat{+} \mathbf{V}$ . Therefore,

$$-\{(-\mathbf{V}) \hat{+} \mathbf{U}\} = (-\mathbf{U}) \hat{+} \mathbf{V} \tag{1.1}$$

If $\hat{+}$ is non-commutative

$$-\{(-\mathbf{V}) \hat{+} \mathbf{U}\} \neq \mathbf{V} \hat{+} (-\mathbf{U}) \tag{1.2}$$

## 2. Lorentz-Einstein Transformation

If a body is traveling with velocity **U** and an observer is traveling with velocity **V,** the relative velocity, $\mathbf{W}_{L-E}$, according to Lorentz-Einstein transformation is given by.

$$\mathbf{W}_{L-E} = (-\mathbf{V}) \vec{+} \mathbf{U} = \frac{\sqrt{1-(\mathbf{V}/c)^2}\,\mathbf{U} + \left\{\left[1-\sqrt{1-(\mathbf{V}/c)^2}\right]\frac{\mathbf{U.V}}{\mathbf{V}^2} - 1\right\}\mathbf{V}}{1-\frac{\mathbf{V.U}}{c^2}} \tag{2.1}$$

We have used the symbol $\vec{+}$ to represent Lorentz-Einstein addition of velocities. If the object and the observer swap positions so that **U** is the velocity of the observer and **V** is the velocity of the object, the relative velocity, $\widetilde{\mathbf{W}}_{L-E}$, is given by

$$\widetilde{\mathbf{W}}_{L-E} = (\mathbf{-U}) \vec{+} \mathbf{V} = \frac{\sqrt{1-(\mathbf{U}/c)^2}\,\mathbf{V} + \left\{\left[1-\sqrt{1-(\mathbf{U}/c)^2}\right]\frac{\mathbf{U.V}}{\mathbf{U}^2} - 1\right\}\mathbf{U}}{1-\frac{\mathbf{V.U}}{c^2}} \tag{2.2}$$

In general they are not equal and opposite.

$$\widetilde{\mathbf{W}}_{L-E} \neq -\mathbf{W}_{L-E} \tag{2.3}$$

An exception is when **U** and **V** are parallel. In that case they are equal.

(2.1) and (2.2) show that

$$\mathbf{V} \vec{+} (-\mathbf{U}) = -\mathbf{W}_{l-E} \neq \widetilde{\mathbf{W}}_{L-E} \tag{2.4}$$

### 3. Non-Associativity of Lorentz Transformation

Consider 3 vectors **U, V** and **W** using definition (2.1) of $\vec{+}$, we can calculate $(\mathbf{U} \vec{+} \mathbf{V}) \vec{+} \mathbf{W}$ and $\mathbf{U} \vec{+} (\mathbf{V} \vec{+} \mathbf{W})$ and show that in general

$$(\mathbf{U} \vec{+} \mathbf{V}) \vec{+} \mathbf{W} \neq \mathbf{U} \vec{+} (\mathbf{V} \vec{+} \mathbf{W}) \tag{3.1}$$

We can by pass the cumbersome calculation and prove (3.1) by using the following theorem.

### 4. Theorem:

If operation $\hat{+}$ is associative

$$-(\mathbf{U} \hat{+} \mathbf{V}) = (-\mathbf{V}) \hat{+} (-\mathbf{U}) \tag{4.1a}$$

And if operation $\hat{+}$ is also non-commutative

$$-(\mathbf{U} \hat{+} \mathbf{V}) \neq (-\mathbf{U}) \hat{+} (-\mathbf{V}) \tag{4.1b}$$

(4.1a) is a necessary condition for associativity. (4.1b) is a necessary condition for associativity and non-commutativity.

*Proof:*

Consider

$$\{\mathbf{U} \hat{+} \mathbf{V}\} \hat{+} (-\mathbf{V}) \hat{+} (-\mathbf{U}) \tag{4.2}$$

Using associativity and $\mathbf{V} \hat{+} (-\mathbf{V}) = 0$ (4.2) becomes

$$\{\mathbf{U} \hat{+} \mathbf{V}\} \hat{+} (-\mathbf{V}) \hat{+} (-\mathbf{U}) = \mathbf{U} \hat{+} \{\mathbf{V} \hat{+} (-\mathbf{V})\} \hat{+} (-\mathbf{U}) = 0 \tag{4.3}$$

Therefore,

$$\{\mathbf{U} \hat{+} \mathbf{V}\} \hat{+} \{(-\mathbf{V}) \hat{+} (-\mathbf{U})\} = 0 \tag{4.5}$$

Comparing (4.5) with

$$(\mathbf{U} \hat{+} \mathbf{V}) \hat{+} -(\mathbf{U} \hat{+} \mathbf{V}) = 0 \tag{4.6}$$

We find that

$$-(\mathbf{U} \hat{+} \mathbf{V}) = (-\mathbf{V}) \hat{+} (-\mathbf{U}) \tag{4.7}$$

If $\hat{+}$ is also non-commutative

$$-\{\mathbf{V} \hat{+} \mathbf{U}\} \neq (-\mathbf{V}) \hat{+} (-\mathbf{U}) \tag{4.8}$$

## 5. Non-Associativity of Lorentz Transformation

(2.4) contradicts (4.7) and (4.8). Therefore, Lorentz transformation is not associative.

## 6. Associativity of Reciprocal Symmetric Transformation

Relative velocity according to reciprocal transformation is defined as

$$\mathbf{W}_{RS} = (\mathbf{-V}) \hat{+} \mathbf{U} = \frac{-\mathbf{V} + \mathbf{U} - i\frac{1}{c}\mathbf{VxU}}{1 - \frac{\mathbf{V.U}}{c^2}} \tag{6.1}$$

A direct calculation will show that for any **U**, **V** and **W**

$$(\mathbf{U} \hat{+} \mathbf{V}) \hat{+} \mathbf{W} = \mathbf{U} \hat{+} (\mathbf{V} \hat{+} \mathbf{W}) \tag{6.2}$$

(6.2) can also be seen from the fact that (6.1) is built from Pauli quaternion[1] and Pauli quaternion is associative.
Definition (6.1) fulfills requirement (4.1)

$$-(\mathbf{V} \hat{+} \mathbf{U}) = (-\mathbf{U}) \hat{+} (-\mathbf{V}) \tag{6.3}$$

## 7. Comparison between Lorentz and Reciprocal Symmetric Transformations

Lorentz transformation and reciprocal symmetric transformations have the same magnitude.

$$|\mathbf{V} \hat{+} \mathbf{U}| = |\mathbf{U}) \mp \mathbf{V}| \tag{7.1}$$

Reciprocal symmetric transformation has a complex x-product term. This x-product term ensures (6.3). In Lorentz transformation (2.1) the product term comes as a dot-product, which does not change sing when order is changed. We have already seen[2] the complex part explicitly shows the rotation hidden in Lorentz transformation. We have also seen[3] that the complex part explicitly shows the origin of spin.

## 8. Conclusion

We have shown that Lorentz Transformation is not associative and that Reciprocal Symmetric transformation is associative. Reciprocal Symmetric transformation fulfills requirement of Lorentz invariance[4]. It is complex. This complex part gives rise to spin[5].

[1] Mushfiq Ahmad, M. Shah Alam, M.O.G. Talukder. Comparison between Spin and Rotation Properties Of Lorentz Einstein and Reflection Symmetric Transformations.http://www.arxiv.org/abs/math-ph/0701067

[2] Mushfiq Ahmad, M. Shah Alam, M.O.G. Talukder. Comparison between Spin and Rotation Properties Of Lorentz Einstein and Reflection Symmetric Transformations.http://www.arxiv.org/abs/math-ph/0701067

[3] Mushfiq Ahmad. Reciprocal Symmetry and the Origin of Spin. http://www.arxiv.org/abs/math-ph/0702043

[4] Mushfiq AhmadReciprocal Symmetry and Equivalence between Relativistic and Quantum Mechanical Concepts. http://www.arxiv.org/abs/math-ph/0611024.

[5] Mushfiq Ahmad. Reciprocal Symmetry and the Origin of Spin. http://www.arxiv.org/abs/math-ph/0702043